\documentstyle[epsfig]{mn2e}

\def\be{\begin{equation}}
\def\ee{\end{equation}}
\def\bea{\begin{eqnarray}}
\def\eea{\end{eqnarray}}

\title[]{Sub-GeV flashes in $\gamma-$ray burst afterglows
as probes of underlying bright far-ultraviolet flares}
\author[]{Yizhong Fan$^{1,2,3}$\thanks{Lady Davis Fellow, E-mail: yzfan@pmo.ac.cn} and Tsvi Piran$^1$
\thanks{tsvi@phys.huji.ac.il}\\
$^1${\sl The Racah Inst. of Physics, Hebrew University, Jerusalem 91904, Israel}\\
$^2$ {\sl Purple Mountain Observatory, Chinese Academy of
Science, Nanjing 210008, China}\\
$^3${\sl National Astronomical Observatories, Chinese Academy of
Sciences, Beijing 100012, China}\\}
\date{Accepted ......
Received ......; in original form ......}

\begin{document}

\maketitle
\begin{abstract}
Bright optical and X-ray flares have been observed in many
Gamma-ray Burst (GRB) afterglows. These flares have been
attributed to  late activity of the central engine. In most cases
the peak energy is not known and it is possible and even likely
that there is a significant far-ultraviolet component. These
far-ultraviolet photons escape our detection because they are
absorbed by the neutral hydrogen before reaching Earth. However,
these photons cross the blast wave produced by the ejecta that
have powered the initial GRB. They can be inverse Compton
upscattered by hot electrons within this blast wave. This process
will produce a strong sub-GeV flare that follows the high energy
(soft X-ray) tail of the far-UV flare but lasts much longer and
can be detected by the upcoming {\em Gamma-Ray Large Area
Telescope} (GLAST) satellite. This signature can be used to probe
the spectrum of the underlying far-ultraviolet flare. The extra
cooling produced by this inverse Compton process can lower the
X-ray emissivity of the forward shock and explain the unexpected
low early X-ray flux seen in many GRBs.

\end{abstract}


\section{Introduction}
\label{sec:MevGevInt}

The XRT on board of {\it Swift} detected, during the last year,
numerous X-ray flares in GRB afterglows (Burrows et al. 2005;
Nousek et al. 2006; Goad et al. 2006; Romano et al. 2006; Falcone
et al. 2006; O'Brien et al. 2006). These observations confirmed
earlier findings of BeppoSAX (Piro et al. 1998, 2005; Galli \&
Piro 2006; in't Zand et al. 2004) and ASCA (Yoshida et al., 1999).
These flares have been interpreted as arising from late time
activity of the central engine (King et al. 2005; Perna et al.
2006 and Proga \& Zhang 2006) producing either internal shocks
(Fan \& Wei 2005; Zhang et al. 2006; Zou, Xu \& Dai 2006; Wu et
al. 2006) or internal magnetic dissipation (Fan, Zhang \& Proga
2005a).

The flare detected in the afterglow of GRB 050502b peaks in the
soft X-rays (Falcone et al. 2006). However, the peak energy of
most flares is unknown (B. Zhang, 2006, private communication). It
is possible, and even likely, that a significant fraction of the
energy or even most of it is emitted in the far-ultraviolet (FUV)
band. For example in the internal energy dissipation model the
typical synchrotron radiation frequency depends sensitively on the
physical parameters (Fan \& Wei 2005; Zhang et al. 2006; Fan et
al. 2005a) and the synchrotron self-absorption frequency is
$\sim{\rm 10^{15}~Hz}$ (Fan \& Wei 2005). It is possible,
therefore, that the observed X-ray flares are the high energy
tails of FUV flares. It is also possible that there are FUV flares
that have not been detected at all. Further support for this idea
arises from the possible interpretation of the optical flare seen
in GRB 050904 (B\"oer et al. 2006) as a late time activity of the
inner engine (Wei, Yan \& Fan 2006).

Even if FUV flares exist they won't be observed as the FUV photons
are absorbed by the neutral hydrogen in the GRB host galaxy as
well as in our Galaxy.  We show here that an underlying FUV flare
will be upscattered and produce (after inverse Compton) a sub-GeV
flare that may be detected by the upcoming {\em Gamma-Ray Large
Area Telescope} (GLAST; see http://glast.gsfc.nasa.gov/)
satellite. Though (sub-)GeV flashes in GRB afterglows could arise
in other scenarios (e.g., M\'esz\'aros \& Rees 1994; Plaga 1995;
Granot \& Guetta 2003; Dermer \& Atoyan 2004; Beleborodov 2005;
Fan, Zhang \& Wei 2005b), the high energy photon flashes predicted
in this Letter can be distinguished easily since they follows the
high energy (soft X-ray) tail of the FUV flares (see also Wang, Li
\& M\'esz\'aros 2006), which is unexpected in any alternative
model. Our prediction could be tested by the cooperation of {\it
Swift} and GLAST. This is possible as {\it Swift} XRT usually
slews to the GRB source in $\sim 100$ seconds and, the field of
view of the GLAST burst monitor (GBM) is all sky not occulted by
the earth and the Large Area Telescope (LAT) will slew to the GRB
direction automatically in $\sim 5$ minutes.

\section{The physical model}\label{sec:MG_model}
\begin{figure}
\begin{picture}(0,160)
\put(0,0){\includegraphics{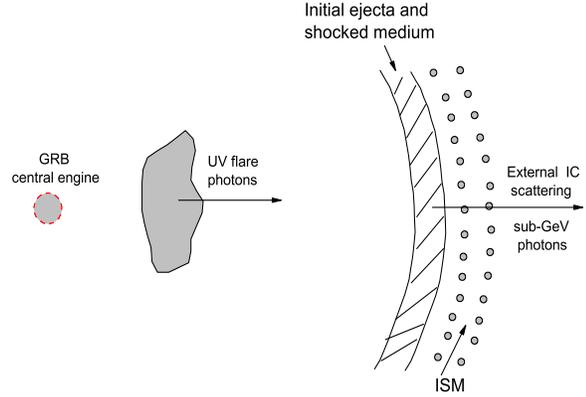}}
\end{picture}
\caption{A schematic cartoon of the flare photons---external
forward shock electrons interaction.} \label{fig:Cartoon}
\end{figure}

The physical process is as follows: A few minutes after the end of
the prompt $\gamma-$ray emission, the central engine becomes
active again and this renewed activity gives rise (either via
internal shocks or via magnetic dissipation) to a strong FUV
flare. At this time the original ejecta that has produced the GRB
has propagated into the circum-burst matter. The blast waves
obtained a Blandford-McKee profile with a strong shock wave in its
front (see Fig. \ref{fig:Cartoon}). The flare FUV photons cannot
be detected on earth as they are absorbed by the neutral Hydrogen.
However, its high energy (soft X-ray) tail might be seen. As the
FUV photons catch up with the blast wave they cool the shock
heated electrons through external inverse Compoton (EIC)
scattering. A very small fraction of the FUV photons is boosted to
a much higher frequency, typically in sub-GeV range. This can be
observed as a sub-GeV flare.

The basic mechanism is radiation produced inside the ejecta at a
small radius is Comptonized in the external blast wave at a much
larger radius. Beloborodov (2005) considered, first, this effect
for Comptonization of prompt 100 keV photons in the reverse shock
of the blast wave. The Comptonization of the prompt photons in the
forward shock was investigated by Fan et al. (2005b). Wang et al.
(2006) considered a scenario in which X-ray photons from X-ray
flares are upscattered to GeV and higher energies. However, as we
discuss later the observed X-ray flare fluxes may be too low to
produce a significant observable GeV signal. FUV flares, that we
discuss here, are motivated by the very soft spectrum of the
observed X-ray flares. Their fluence is not constrained by current
observations and the sub-GeV flare that we predict is the only way
known to explore their existence. Furthermore, as we show below, a 
FUV fluence comparable to the fluence of current X-ray flares will
produce a signal that can be observed by GLAST

We focus on the FUV flares taking place at $100-1000$ seconds
after the burst since (i) $\gamma_m$ decreases rapidly with time.
At later times, the scattered photons are not in the high energy
range. (ii) The early FUV flares are relatively energetic and
contain more seed photons.

For our purpose the total number of seed photons (rather than the
total energy) is more important  as this determine the total
number of high energy (upscattered) photons. To obtain a reliable
estimate of the number of detected photons we need to calculate
the probability of upscattered of a seed photon by the forward
shock electrons, i.e., the optical depth  of these electrons
(\S\ref{sec:depth}). We also need to estimate the number of the
flare photons (\S\ref{sec:photon}). A simple estimate of these
factors yields, for an ISM surrounding:
\begin{eqnarray*}
 N_{\rm obs} &\approx& \sigma_T n_0 (R/3) {{\cal F} \over h \nu_{uv}} S\\
   &\approx& 10 n_0 R_{17} {\cal F}_{-6} (\nu_{uv}/0.01 {\rm keV})^{-1}
(S_{\rm _{GLAST}}/8000 {\rm cm}^2),
\end{eqnarray*}
where  $R$ is the radius of the forward shock, $n_0$ the
surrounding density, ${\cal F}$ is the fluence of the seed FUV
photons' flare and $\nu_{uv}$ is their typical frequency. $S$ is
the detectors area and  $S_{\rm _{GLAST}}\sim 8000{\rm cm^2}$ is
the effective area of Large Area Telescope (LAT) onboard GLAST .
Here and throughout this text, the convention $Q_x=Q/10^x$ has
been adopted in cgs units. Later we elaborate on these estimates.
We also need to take into account the important correction caused
by the anisotropic emission of the scattered photons (in the
comoving frame of the shocked material), which has been ignored
previously (see eqs. (\ref{eq:j}-\ref{eq:f_cor}) for details).

\subsection{The optical depth}\label{sec:depth}
In the standard GRB afterglow model (e.g. Sari, Piran \& Narayan
1998; Piran 1999) the blast wave propagates into a constant
density ISM. In this case
\begin{equation}
\gamma_m\simeq 1.7\times 10^3~\epsilon_{e,-1} C_p
E_{k,53}^{1/8}n_0^{-1/8}t_3^{-3/8}[(1+z)/2]^{3/8},
\end{equation}
where $E_k$ is the isotropic kinetic energy of the initial GRB
outflow, $C_p\equiv 13(p-2)/[3(p-1)]$, $p\sim 2.3$ is the
power-law index of the shocked electrons, $t$ is the observer's
timescale in units of second, $z$ is the redshift of the burst,
$\epsilon_e$ and $\epsilon_B$ represent the fractions of shock
energy given to the electrons and magnetic filed, respectively.
The  Lorentz factor of the cooling electrons is
\begin{equation}
\gamma_c \simeq 1.5\times 10^4 ~E_{k,53}^{-{3\over
8}}\epsilon_{B,-2}^{-1}n_0^{-{5\over 8}} t_3^{1\over
8}[(1+z)/2]^{-{1\over8}}(1+Y)^{-1},
\end{equation}
where $Y$ is the Compton parameter, which is a sum of the
synchrotron self-Compton (SSC) parameter $Y_{_{SSC}}$ and the EIC
parameter $Y_{_{EIC}}$. Following Fan \& Piran (2006; see Appendix
A), $Y_{_{SSC}}\simeq
\{-(1+Y_{_{EIC}})+\sqrt{(1+Y_{_{EIC}})^2+4\eta
\eta_{_{KN}}\epsilon_e/\epsilon_B}\}/2$, where
$\eta=\min\{1,(\gamma_m/\gamma_c)^{(p-2)}\}$ (e.g. Sari, Narayan
\& Piran 1996). The coefficient  $0\leq \eta_{_{KN}}\leq 1$
accounts for the Klein-Nishina effect. For $\bar{\gamma}_e=\min \{
\gamma_c, \gamma_m\}\sim 10^3$, $\eta_{_{KN}}\sim 1$  (see
Appendix B. of Fan \& Piran 2006). To get $\gamma_c$, we also need
to estimate the parameter $Y_{_{EIC}}$.

In the shock front, the magnetic energy density can be estimated
as
\begin{equation}
U_B\simeq 0.1~{\rm ergs
~cm^{-3}}~\epsilon_{B,-2}E_{k,53}^{1/4}n_0^{3/4}t_3^{-3/4}[(1+z)/2]^{3/4}.
\end{equation}
At $t$, the forward shock front reaches the radius $R\simeq
1.9\times 10^{17}~{\rm
cm}~E_{k,53}^{1/4}n_0^{-1/4}t_3^{1/4}[(1+z)/2]^{-1/4}$. In the
forward shock region, the  energy density of FUV photons of the
flare can be estimated as
\begin{eqnarray}
U_{ph} &\simeq & {L_{ph} \over 4\pi R^2 \Gamma^2 c
}\nonumber\\
&=& 0.4~{\rm ergs ~cm^{-3}}~L_{ph,49}E_{k,53}^{-3/4}n_0^{3\over
4}t_3^{1 \over 4}({1+z \over 2})^{-1/4},
\end{eqnarray}
where $L_{ph}$ is the luminosity of the flare. Note that the flare
lasts $\Delta T(<t)$. Its contribution to the cooling of the
electrons can be estimated by the EIC parameter
\begin{eqnarray}
Y_{_{EIC}} &\approx & (\Delta T/t)(U_{ph}/U_B) \nonumber\\
& \simeq & 4 (\Delta T/t)L_{ph,49}
\epsilon_{B,-2}^{-1}E_{k,53}^{-1}t_3[(1+z)/2]^{-1}.
\label{eq:Y_EIC}
\end{eqnarray}

Provided that $\Delta T/t\sim 0.3$, $Y=Y_{_{SSC}}+Y_{_{EIC}}\simeq
3$ for typical parameters and $\gamma_c\sim 4\times 10^3$. We have
$\bar{\gamma}_e=\min \{\gamma_m, \gamma_c\} \sim 2\times 10^3$.
These electrons scatter on the flare photons and boost them to the
typical energy (in the observer frame)
\begin{equation}
h\nu_{\rm obs}\sim 2\bar{\gamma}_e^2 h\nu_{\rm uv}= 80~{\rm MeV}~
\bar{\gamma}_{e,3.3}^2(h\nu_{\rm uv}/0.01 {\rm keV}),
\end{equation}
where $h$ is Plank's constant and $\nu_{\rm uv}$ is the typical
initial photon frequency.

Obviously, the scattering is in the Thompson regime. The optical
depth for a flare photon to be scattered can be estimated by
\begin{equation}
\tau_{\rm ISM} \simeq \sigma_T n R/3\simeq 4.2\times
10^{-8}~E_{k,53}^{1/4}n_0^{3/4}t_3^{1/4}[(1+z)/2]^{-1/4},
\end{equation}
where $\sigma_T$ is the Thompson cross section.

In the stellar wind model (Dai \& Lu 1998; M\'esz\'aros, Rees \&
Wijers 1998), $n=3\times 10^{35}A_* R^{-2}~{\rm cm^{-3}}$, where
$A_*=[\dot{M}/10^{-5}M_\odot~{\rm yr^{-1}}][v_w/(10^8{\rm cm}~{\rm
s^{-1}})]$ (Chevalier \& Li 2000), $\dot{M}$ is the mass loss rate
of the progenitor, $v_w$ is the velocity of the wind. Now
$\gamma_m\simeq 970 \epsilon_{e,-1} C_p
E_{k,53}^{1/4}A_*^{-1/4}t_3^{-1/4}[(1+z)/2]^{1/4}$ and
$\gamma_c\simeq
280\epsilon_{B,-2}^{-1}E_{k,53}^{1/4}A_*^{-5/4}t_3^{3/4}[(1+z)/2]^{-3/4}(1+Y)^{-1}$.
Similar to the ISM case, at $t$ the forward shock front reaches
the radius $R \simeq  2.7\times 10^{16}~{\rm
cm}~E_{k,53}^{1/2}A_*^{-1/2}t_3^{1/2}[(1+z)/2]^{-1/2}$, $U_B
\simeq  13~{\rm ergs
~cm^{-3}}~\epsilon_{B,-2}E_{k,53}^{-1/2}A_*^{3/2}t_3^{-3/2}({1+z
\over 2})^{3/2}$, $U_{ph} = 87~{\rm ergs
~cm^{-3}}~L_{ph,49}E_{k,53}^{-3/2}A_*^{3\over 2}t_3^{-1 \over
2}({1+z \over 2})^{1/2}$, and the EIC parameter can be estimated
by
\begin{equation}
Y_{_{EIC}}\simeq 6.7(\Delta T/t) \epsilon_{B,-2}^{-1} L_{ph,49}
E_{k,53}^{-1} t_3({1+z\over 2})^{-1}.
\end{equation}
Provided that $\Delta T/t\sim 0.3$, we have
$Y=Y_{_{SSC}}+Y_{_{EIC}}\simeq 3.4$ and $\gamma_c\sim 70$. So
$\bar{\gamma}_e\sim 70$ for $t\sim 10^3$ s. The possibility of one
flare photon being scattered (i.e., the optical depth) can be
estimated by
\begin{equation}
\tau_{\rm wind} \simeq 7.3 \times
10^{-6}~A_*^{3/2}E_{k,53}^{-1/2}t_3^{-1/2}[(1+z)/2]^{1/2}.
\end{equation}
Now most scattered photons are in the sub MeV band and the count
number is $\sim 0.05$ cm$^{-2}$ (where eq. (\ref{eq:N_tot}) and
eq.(\ref{eq:f_cor}) have been taken into account), which is
undetectable for the {\it Swift} BAT. The tens MeV photons
(resulting in the keV flare photons-forward shock electrons
interaction) may be still detectable for the GLAST. But the counts
rate is not higher than that of the ISM case. On the other hand,
there are just a small fraction of bursts were born in the stellar
wind (e.g., Chevalier \& Li 2000; Panaitescu \& Kumar 2002). So we
do not discuss this case further.

\subsection{The duration of the high energy emission}\label{sec:duration}
As pointed out in Beloborodov (2005), the upscattered photons are
de-collimated and their arrival time is affected by the spherical
curvature of the blast wave. The duration of the high energy
emission thus can be estimated as
\begin{equation}
T \sim \Delta T +(1+z)R/2\Gamma^2,
\end{equation}
we have $T\sim 4t$ in the ISM case and $T\sim 2t$ in the wind
case, which could be much longer than $\Delta T$. The duration
increases when the anisotropic radiation of the up-scattered
photons (see eq. [\ref{eq:j1}]) has been taken into account
because now the strongest emission are from $\theta \sim
1/\Gamma$.

As a result, most of the up-scattered photons will arrive after
the FUV flare. This lagging behavior is a signature of
upscattering of internal radiation in the external blast wave,
which may be tested by the observations.

\subsection{The total number of soft photons}\label{sec:photon}
Assuming that the spectrum of the flare has the form
$F_{\nu}\propto \nu^{-\beta_{_{XRT}}}~{\rm for~\nu>\nu_{\rm uv}}$,
where $\beta_{_{XRT}}\sim 1.2$, as reported in most X-ray flares
(e.g. O'Brien et al. 2006).  The total number of soft FUV photons
reaching us (in unit area and without absorption) can be estimated
by
\begin{eqnarray}
N_{\rm tot} &\simeq & {\beta_{_{XRT}}-1 \over
\beta_{_{XRT}}}{{\cal F} \over h\nu_{\rm uv}}, \label{eq:N_tot}
\end{eqnarray}
where ${\cal F}$ is the energy fluence of the flare.

\subsection{The detectability of sub-GeV photons}
The interaction between the photon beam and the isotropic
relativistic electrons (i.e., the anisotropic IC scattering) has
been discussed externsively (Brunetti 2001 and the references
therein). Here the scattering is in the Thompson regime and the
electrons are ultra-relativistic (their distribution is
$n(\gamma_e)=K_e \gamma_e^{-\delta}$ for
$\min\{\gamma_c,\gamma_m\}<\gamma_e<\max\{\gamma_c,\gamma_m\}$),
the emissivity can be approximated as (i.e., eq. (43) of Brunetti
2001)
\begin{eqnarray}
j(\cos \theta_{s},\nu_s)&=& K_e r_0^2c{(1-\cos \theta_s)^{(\delta
+1)/2}(\delta^2+4\delta+11)\over (\delta+1)
(\delta+3) (\delta+5)} \nonumber\\
&& \nu_s^{-(\delta-1)/2}\int \nu^{(\delta-1)/2}n(\nu)d\nu,
\label{eq:j}
\end{eqnarray}
where $n(\nu)$ is the energy distribution of the seed photons,
$r_0$ is the classic radius of the electron, $\theta_s$, is the
scattering angle (measured in the comoving frame of the shocked
material and set to zero on the line of the velocity vector),
$\nu_s$ is the frequency of the scattered photon (measured in the
comoving frame of the shocked material). As shown in eq.
(\ref{eq:j}) the scattered power has a maximum at $\theta_s=\pi$
and goes to zero for small scattering angles. The shocked medium
moves toward us with a bulk Lorentz factor about tens. The photons
scattered in the comoving frame at an angle $\theta_s\sim \pi/2$
from the velocity vector are those making an angle $\theta \sim
1/\Gamma$ with the line of sight in the observer frame  $\cos
\theta_s=(\cos\theta-\beta)/(1-\beta \cos \theta)$. So the
received power is depressed (relative to the isotropic seed photon
case) but not significantly, as shown below.

If the seed photons are also isotropic (so are the scattered
ones), integrating eq. (\ref{eq:j}) yields the well known result
(e.g. Blumenthal \& Gould 1970)
\begin{eqnarray}
j(\nu_s)&=& \pi r_0^2 K_e c 2^{\delta
+3}{(\delta^2+4\delta+11)\over (\delta+1)
(\delta+3)^2 (\delta+5)} \nonumber\\
&& \nu_s^{-(\delta-1)/2}\int \nu^{(\delta-1)/2}n(\nu)d\nu.
\label{eq:j1}
\end{eqnarray}
What we care about is the divergency of receiving number of
scattered photons at $\nu_{\rm obs}={\cal D}^{-1}\nu_s$ between
the photon beam case and the isotropic photon case (${\cal
D}=\Gamma (1-\beta \cos \theta)$ is the Doppler factor), which is
represented by $f_{\rm cor}$ and can be estimated as (e.g. Rybicki
\& Lightman 1979)
\begin{equation}
f_{\rm cor} \simeq {\int_0^{\theta_j} j(\cos \theta_{s},{\cal
D}\nu_{\rm obs}){\cal D}^{-3} \sin \theta d\theta \over
\int_0^{\theta_j} j({\cal D}\nu_{\rm obs}){\cal D}^{-3} \sin
\theta d\theta}, \label{eq:f_cor}
\end{equation}
where $\theta_j$ is the jet half-opening angle of the ejecta. We
have $f_{\rm cor}\simeq 0.4$ for $\delta \sim 2.3$ and
$\theta_j\gg 1/\Gamma$.

In the ISM case,
\begin{eqnarray}
 N_{\rm obs} &\sim & f_{\rm cor}\tau_{\rm ISM} N_{\rm tot}S_{\rm _{GLAST}} \nonumber\\
&=& 1.3C_{\beta}{\cal F}_{-6}({h\nu_{\rm uv} \over 0.01{\rm keV}
})^{-1} E_{k,53}^{1/4}n_0^{3\over 4}t_3^{1\over 4}({1+z \over
2})^{-1\over 4}, \label{eq:n_ISM}
\end{eqnarray}
sub-GeV photons can be collected by GLAST, where $C_\beta \equiv
6(1-\beta_{_{XRT}})/\beta_{_{XRT}}$. Usually at least five photos
are needed to claim  a detection (Zhang \& M\'esz\'aros 2001), so
we need $n\sim 10~{\rm cm^{-3}}$, which is typical (Panaitescu \&
Kumar 2002).

The effective area of EGRET onboard Compton Gamma Ray Observatory
(CGRO) is $S_{\rm _{EGRET}}\sim 1500~{\rm cm^2}$. A rather high
circumburst density of $n\sim 100~{\rm cm^{-3}}$ is needed  to get
5 sub-GeV photons. Afterglow modeling (Panaitescu $\&$ Kumar 2002)
suggests that such a high density is uncommon around GRB
progenitors. They may be the reasons for the rare detections of
delayed sub-GeV photon flashes by EGRET (see \S\ref{sec:MG_case}
for details).

Before turning to a comparison with observations we ask two
questions. First we ask whether SSC process of the electrons
accounting for the FUV flares can produce sub-GeV photons. We then
ask what are the implications of the cooling due to the IC process
on the forward shock emission.

The answer to the first question, can these sub-GeV photons be
attributed to the SSC radiation of the electrons accounting for
the FUV flares is very likely negative. Firstly, the outflow
powering the FUV flares may be highly magnetized (Usov 1992;
Thompson 1994; Lyutikov \& Blandford 2003; Spruit, Daigne \&
Drenkhahn 2001; Fan et al. 2005a; Proga \& Zhang 2006) in which
case the synchrotron self-Compton radiation is too weak to be
detectable. Secondly, if the late baryonic internal shock emission
peaks in the FUV band (i.e., $\nu_{\rm uv}$), the typical SSC
frequency should be $\sim \gamma_{e,m}^2 \nu_{\rm uv}\sim 100
~{\rm keV}~\gamma_{e,m,2}^2(\nu_{\rm uv}/0.01{\rm keV})\ll {\rm
tens~MeV}$, where $\gamma_{e,m}\sim 100$ is the minimum Lorentz
factor of electrons accelerated in the late internal shocks (see
also Wei et al. 2006). Its contribution to sub-GeV emission flux
is unimportant. If the typical synchrotron radiation frequency of
late internal shocks is in X-ray band, the SSC radiation may peak
in tens MeV band (Wang et al. 2006). However, as we have already
mentioned in section 1, for most ``X-ray flares" detected so far,
the peak energy may be lower than 0.2 keV. So the tens MeV
emission from the SSC process may be infrequent.

We need to verify that the sub-GeV photons won't be absorbed by
the high energy tail of the FUV flare photons. The pair production
optical depth for photons with energy $\sim 1$ GeV (absorbed by
the flare photons with energy $\epsilon_{a,{\rm obs}}\sim 2(\Gamma
m_e c^2)^2/[(1+z)^2 {\rm GeV}]\sim 0.2 {\rm
MeV}~E_{k,53}^{1/4}n_0^{-1/4}t_3^{-3/4}[(1+z)/2]^{-5/4}$) can be
estimated as (e.g., Svensson 1987)
\begin{eqnarray}
\tau_{\gamma \gamma} ({\rm 1 GeV})\simeq {11 \sigma_T
N_{>\epsilon_{a,{\rm obs}}} \over 720 \pi R^2} \sim 10^{-5},
\end{eqnarray}
where $N_{>\epsilon_{a,{\rm obs}}}= {\beta_{_{ XRT}}-1 \over
\beta_{_{ XRT}}} ({ h\nu_{\rm uv} \over \epsilon_{a,{\rm obs}}
})^{\beta_{_{XRT}}}{4\pi D_L^2{\cal F}\over (1+z)^2 h\nu_{\rm
uv}}$ is the total flare photon number satisfying $h
\nu>\epsilon_{a,{\rm obs}}$. Clearly such a small optical depth
won't affect the sub-GeV flux.

FUV flares may play  an additional role. Consider the possibility
that after the cease of the $\gamma-ray$ burst, the central engine
does not turn off and gives rise to long term but sharply decaying
soft radiation component (mainly in far-ultraviolet band). The IC
process of these FUV photons cools the  forward shock electrons
and the IC parameter $Y$ may be dominated by $Y_{_{EIC}}$. This
will reduce the early X-ray flux emitted by these electrons since
the X-ray flux recorded by XRT is $\propto (1+Y)^{-1}$ (e.g., eq.
(6) of Fan \& Piran 2006).  For illustration, with $L_{ph}\sim
6\times 10^{49}~{\rm ergs}~\epsilon_{B,-2}
E_{k,53}(t/400)^{-1.7}[(1+z)/2]^{1.7}$ for $400 {\rm s}<t<10^4{\rm
s}$, we have $Y_{_{EIC}}\approx 10 (t/400)^{-0.7}$ (we have used
Eq. [\ref{eq:Y_EIC}] with $\Delta T/t=1$ and with typical
parameters). Depending on $Y_{_{SSC}}$ this reduces the X-ray flux
by a factor of 3-10 and results in a slow declines as $t^{-0.5}$
rather than as $t^{-1.2}$. This provides a possible explanation to
the puzzle of weak slowly declining X-ray flux observed by {\it
Swift} in many GRBs (Nousek et al. 2006). Note, however, that this
process requires a significant $\sim 10^{52}$ergs FUV emission.

\section{Possible candidates of the predicted sub-GeV flashes}\label{sec:MG_case}
EGRET has detected more than 30 GRBs with  sub-GeV photon emission
(e.g., Schneid et al. 1992, 1995; Sommer et al. 1994; Hurley et
al. 1994; Schaefer et al. 1998; Gonz\'alez et al. 2003). In some
events, which interest us here, the duration of the sub-GeV
emission is longer than that of keV-MeV emission.

{\bf GRB 930131:} The keV-MeV emission lasted $\sim 50$ s but two
$\sim 100$ MeV  photons  were detected at $74$ s and $99$ s after
the BATSE trigger (Sommer et al. 1994). Both the energy of photons
and the count number seem to be consistent with the predictions of
the far ultraviolet flare-forward shock interaction model.

{\bf GRB 940217:} The sub-GeV emission lasted more than 5000
seconds and it included also a 18 GeV photon (Hurley et al. 1994).
The spectrum in the energy range $1~{\rm MeV}$ to 18 GeV,  cannot
be fitted with a simple power law (see Fig. 3 of Hurley et al.
1994). A new spectral component in the energy range larger than
several tens MeV is needed. Possible models include the
interaction of ultra-relativistic protons with a dense cloud (Katz
1994), SSC scattering in early forward and reverse shocks
(M\'esz\'aros \& Rees 1994) and an electromagnetic cascade of TeV
$\gamma-$rays in the infrared/microwave background (Plaga 1995).

The long term strong X-ray flare detected in the high redshift
burst GRB 050904 (e.g. Watson et al. 2006) hints that the long
term sub-GeV emission of GRB 940217 may be explained as
upscattering of FUV flare by the forward shock electrons (see also
Wang et al. 2006). Of course we cannot tell now if there was an
underlying long term FUV in GRB 940217, but its existence in other
bursts makes this a viable model.

{\bf GRB 941017} has shown in addition to the typical GRB emission
a distinct high-energy spectral component extending from $<$ a few
MeV to $>$ 200 MeV. The high-energy component carried at least 3
times more energy than the lower energy component. The hard high
energy component lasted 200 seconds, much longer than the low
energy $\gamma-$ray emission that lasted $\sim 77$ s (Gonz\'alez
et al 2003). While various models have been put forward (Granot \&
Guetta 2003; Pe'er \& Waxman 2004), the observed hard spectrum has
not been well reproduced except in the neutral beam model (Dermer
\& Atoyan 2004) and the prompt $\gamma-$rays---reverse shock
interaction model (Beloborodov 2005). It is also challenging to
explain this rather hard spectrum with the FUV flares--forward
shock interaction model discussed here.

In addition to  GRB 041017, Gonz\'alez et al. (2003) have found
significant sub-GeV emission in other 25 bright GRBs. In these
cases the high energy spectra are consistent with the single power
law component observed by BATSE. This, of course, favors the
internal shock synchrotron radiation model. Therefore,  we are
left in the EGRET era, with only two candidates (e.g., GRB 930131
and GRB 940217) of the predicted sub-GeV photon emission are
available. The upcoming GLAST may be able to detect more events.

\section{Discussion and summary}\label{sec:MG_Dis}
Bright X-ray flares have been detected a large group of {\it
Swift} GRB afterglows. These flares have been attributed to late
activity of the central engine. In most cases the peak energy is
not known and it is possible that there is a significant
far-ultraviolet component. These far-ultraviolet photons escape
our detection because they are absorbed by the neutral hydrogen
both in host galaxy and in our Galaxy  before reaching Earth. We
suggest here that these far-ultraviolet photons are IC upscattered
by hot electrons within the shock that is ahead of them. This
shock is driven by the blast wave produced by the ejecta that
powered the initial GRB. This IC process will produce a strong
sub-GeV burst what can be detected by upcoming {\em Gamma-Ray
Large Area Telescope} (GLAST) satellite if the
far-ultraviolet/X-ray flare is bright enough (the energy fluence
${\cal F}\sim 10^{-6}~{\rm ~erg~cm^{-2}}$). Alternatively, if most
flare photons are in keV band rather than in far-ultraviolet band,
the total number of photons being scattered (and the detected
number of photons) is much smaller though the typical energy is
much higher (Generally, they are in GeV-TeV energy range, see Wang
et al. 2006). It is not easy to collect enough photons for
significant detection.

We have also analyzed the sub-GeV detections of EGRET in view of
this model. In some events (for example, GRB 930131, GRB 940217
and GRB 941017), the sub-GeV photons are delayed. Out of these
bursts, the spectrum of GRB 941017 is rather hard and it cannot be
explained by this model. On the other hand the other two events
seem to be consistent with the model. However, these sub-GeV
photons could be generated in other scenarios (e.g., M\'esz\'aros
\& Rees 1994 Plaga 1995; Dermer \& Atoyan 2004; Fan et al. 2005b).
Without simultaneous soft X-ray observations we cannot confirm our
model.

Finally, we point out that the extra cooling induced by continuous
flux of FUV photons that pass through the forward shock reduces
the X-ray flux produces by this shock front. Thus if the central
engine does not turn off and gives out a  significant emission of
FUV photons several thousand seconds after the GRB, this emission
will reduce the X-ray flux emitted by the forward shock at the
beginning of the afterglow phase. This possibility provides an
alternative explanation to the weak and slowly declining early
X-ray light curve observed by {\it Swift} in many GRBs (Nousek et
al. 2006). As this process involves emission of sub-GeV photons it
could be tested by simultaneous observations of {\it Swift} and
GLAST in the near future.

\section*{Acknowledgments}
Y. Z. Fan thanks Bing Zhang, D. M. Wei and X. Y Wang for comments
and Z. Li., Y. F. Huang and X. F. Wu for kindful help. We thank
the referee for helpful comments. This work is supported by
US-Israel BSF. TP acknowledges the support of Schwartzmann
University Chair. YZF is also supported by the National Natural
Science Foundation (grants 10225314 and 10233010) of China, and
the National 973 Project on Fundamental Researches of China
(NKBRSF G19990754).


\begin{thebibliography}{99}
\bibitem[]{} Beloborodov A. M., 2005, ApJ, 618, L13
\bibitem[]{} Blumenthal G. R., Gould R. J., 1970, Rev. Mod. Phys.,
42, 237
\bibitem[]{} Bo\"er M., Atteia J. L., Damerdji Y., Gendre B., Klotz A.,
Stratta G. 2006, ApJ, 638, L71
\bibitem[]{} Brunetti G., 2001, Astroparticle Phys., 13, 107
\bibitem[]{} Burrows D. N., et al., 2005, Science, 309, 1833
\bibitem[]{} Chevalier R. A., Li Z. Y., 2000, ApJ, 536, 195
\bibitem[]{} Connaughton V., 2002, ApJ, 567, 1028
\bibitem[]{} Dai Z. G., Lu T., 1998, MNRAS, 298, 87
\bibitem[]{} Dermer C. D., Atoyan A., 2004, A\&A, 418, L5
\bibitem[]{} Falcone A. D., et al. 2006, ApJ, accepted (astro-ph/0512615)
\bibitem[]{} Fan Y. Z., Piran T., 2006, MNRAS, in press (astro-ph/0601054)
\bibitem[]{} Fan Y. Z., Wei D. M., 2005, MNRAS, 364, L42
\bibitem[]{} Fan Y. Z., Zhang B., Proga D., 2005a, ApJ, 635, L129
\bibitem[]{} Fan Y. Z., Zhang B., Wei D. M., 2005b, ApJ, 629, 334
\bibitem[]{} Galli A.,  Piro L., 2006, A\&A, submitted (astro-ph/0510852)
\bibitem[]{} Goad M. R., et al., 2006, A\&A, 449, 89
\bibitem[]{} Gonz\'alez M. M., et al., 2003, Nature, 424, 749
\bibitem[]{} Granot J., Guetta D., 2003, ApJ, 598, L11
\bibitem[]{} Hurley K., et al., 1994, Nature, 372, 652
\bibitem[]{} in't Zand J. J. M., Heise J., Kippen R. M., Woods, P. M., Guidorzi, C., Montanari E., Frontera F.,
2004, ASPC, 312, 181
\bibitem[]{} Katz J. I., 1994, ApJ, 432, L27
\bibitem[]{} King A., O'Brien P. T., Goad M. R., Obsorne J.,
Olsson E., Page K., 2005, ApJ, 630, L113
\bibitem[]{} Lyutikov M., Blandford R., 2003 (astro-ph/0312347)
\bibitem[]{} MacFadyen A. I., Woosley S. E.,
Heger A., 2001, ApJ, 550, 410
\bibitem[]{} M\'{e}sz\'{a}ros P., Rees M. J., 1994, MNRAS, 269,
L41
\bibitem[]{} M\'esz\'aros P., Rees M. J., Wijers R. A. M. J.,
 1998, ApJ, 499, 301
\bibitem[]{} Nousek J. A., et al., 2005, ApJ, in press (astro-ph/0508332)
\bibitem[]{} O'Brien P. T., et al., 2006, ApJ, submitted (astro-ph/0601125)
\bibitem[]{} Panaitescu A., Kumar P., 2002, ApJ, 571, 779
\bibitem[]{} Pe'er A., Waxman E., 2004, ApJ, 603, L1
\bibitem[]{} Perna R., Armitage P.J., Zhang B., 2006, ApJ, 636, L29
\bibitem[]{} Plaga R., 1995, Nature, 374, 430
\bibitem[]{} Piran T., 1999, Phys. Rep., 314, 575
\bibitem[]{} Piro L., et al., 1998, A\&A, 331, L41
\bibitem[]{} Piro L., et al., 2005, ApJ, 623, 314
\bibitem[]{} Proga D., Zhang B., 2006, ApJL, submitted (astro-ph/0601272)
\bibitem[]{} Romano P. et al., 2006, A\&A, in press (astro-ph/0601173)
\bibitem[]{} Rybicki G. B., Lightman A. P., Radiative Processes in Astrophysics
(Wiley, New York. 1979)
\bibitem[]{} Sari R., Narayan R., Piran T., 1996, ApJ, 473, 204
\bibitem[]{} Sari R., Piran T., Narayan R. 1998, ApJ, 497, L17
\bibitem[]{} Schaefer B. E., et al. 1998, ApJ, 492, 696
\bibitem[]{} Schneid E. J., et al., 1992, A\&A, 255, L13
\bibitem[]{} Schneid E. J., et al., 1995, ApJ, 453, 95
\bibitem[]{} Sommer M., et al., 1994, ApJ, 422, L63
\bibitem[]{} Spruit, H. C., Daigne, F., Drenkhahn, G., 2001, A\&A,
369, 694
\bibitem[]{} Svensson R., 1987, MNRAS, 227, 403
\bibitem[]{} Thompson C., 1994, MNRAS, 270, 480
\bibitem[]{} Usov V. V. 1992, Nature,  357, 472
\bibitem[]{} Wang X. Y., Li Z.,  M\'esz\'aros P., 2006, ApJL, submitted (astro-ph/0601229)
\bibitem[]{} Waston D., et al., 2006, ApJ, 637, L69
\bibitem[]{} Wei D. M., Yan T., Fan Y. Z., 2006, ApJ, 636, L69
\bibitem[]{} Wu X. F., Dai Z. G., Wang X. Y., Huang Y. F., Feng L. L., Lu T.,
 2006, ApJ, submitted (astro-ph/0512555)
\bibitem[]{} Yoshida A., Namiki C., Otani C., Kawai N., Murakami T.,
Ueda Y., Shibata R., Uno S., 1999, A\&A supp, 138, 433
\bibitem[]{} Zhang B., Fan Y. Z., Dyks J., Kobayashi S., M\'esz\'aros P.,
Burrows D. N., Nousek J. A., Gehrels N.  2006, ApJ, in press
(astro-p/0508321)
\bibitem[]{} Zhang B., M\'esz\'aros P., 2001, ApJ, 559, 110
\bibitem[]{} Zou Y. C., Xu D., \& Dai Z. G., 2006, ApJL, submitted (astro-ph/0511205)

\end{thebibliography}
\end{document}